\documentclass[epj]{svjour}
\usepackage{graphicx} 
\usepackage{latexsym}
\usepackage{amssymb}
\usepackage{textcomp}
\newcommand{\lo}{\left \langle}
\newcommand{\rc}{\right \rangle}

\usepackage{amssymb}
\begin{document}
\title{Bubbles and denaturation in DNA}

\author{Titus S. van Erp \inst{1} \email{Titus.VanErp@biw.kuleuven.be} \and Santiago Cuesta-L\'opez\inst{2,3} \and
Michel Peyrard\inst{2}}

\institute{Centre for Surface Chemistry and Catalysis,
Catholic University of Leuven, Kasteelpark Arenberg 23, 3001 Leuven, Belgium
\and 
Laboratoire de Physique, Ecole Normale Sup\'erieure de Lyon,
46 all\'ee d'Italie, 69364 Lyon Cedex 07, France \and
Dept. Condensed Matter Physics and Institut of Biocomputation and
Complex Systems. University of Zaragoza, c/ Pedro Cerbuna s/n 50009 Spain
}

\abstract{
The local opening of  DNA  is an  intriguing  
phenomenon from a statistical physics 
point of view, but is also essential for its biological function. 
For instance, the transcription and replication of our genetic code
can not take place without 
the unwinding of the  DNA double helix. 
Although these biological processes are  driven by proteins,
there might well be a relation  between these biological openings
and the spontaneous bubble formation due to thermal fluctuations.
Mesoscopic models, like the Peyrard-Bishop-Dauxois model,  have  
fairly accurately  reproduced  some experimental  denaturation curves and 
the sharp phase transition in the thermodynamic limit.
It is, hence, tempting to see whether these models could be used to predict 
the biological activity of DNA. 
In a previous study, we introduced a method that allows to obtain very
accurate results on this subject, which showed that some previous
claims in this direction, based on molecular dynamics studies, were
premature.  This could either imply that the present PBD should be
improved or that biological activity can only be predicted in a more
complex frame work that involves interactions with proteins and super
helical stresses.  In this article, we give detailed description of
the statistical method introduced before.  Moreover, for several DNA
sequences, we give a thorough analysis of the bubble-statistics as
function of position and bubble size and the so-called
$l$-denaturation curves that can be measured experimentally.  These
show that some important experimental observations are missing in the
present model.  We discuss how the present model could be improved.
}


\maketitle
\section{Introduction}
The process of DNA denaturation has intrigued 
both biologists as statistical physicists.
Large openings, the so-called â DNA bubblesâ
are supposed to allow the formation of some specific DNA structures, 
such as the T-loop that stabilizes the end of the chromosomes.
The opening of the DNA double helix is also a mandatory step 
for the transcription and the replication of the genetic code.
In addition, the bonds between bases on opposite strands can break 
due to thermal fluctuations which can occur even at 
room or physiological temperatures. 
These thermally induced DNA bubbles can be  
several base-pairs long and tend to increase 
at higher temperatures,
which eventually results in 
the complete denaturation or the melting of DNA. 
An intriguing question we could ask ourselves 
is how the formation of bubbles depend on the
base-pair specific sequence and how thermally induced bubbles relate 
to biophysical DNA unwinding mechanisms that are involved in the
transcription and replication.  Although these biological
processes are driven by proteins,
the intrinsic fluctuations of DNA itself might play an important role.
Hence, one could even question whether biological active sites 
could be predicted by thermally induced bubbles in 
absence of any proteins~\cite{ChoiNuc2004,KalosEPL,vanerpPRL}.

Experimentally, the thermally induced denaturation can be monitored 
as the 
breaking of the base-pairs is accompanied with a large increase
of UV absorbance near 260 nm. In fact, the UV absorbance measures the reduction of   base-pairing and -stacking 
when the  DNA molecule denaturates. Using this technique, it was found that 
large synthetical fabricated homopolymers denaturate 
suddenly within a very small temperature 
interval~\cite{Inman}. This indicates that the process resembles a true 
first order phase transition. 
On the other hand, natural heterogeneous DNA polymers denaturate in multiple 
steps and the shape of this denaturation curve is highly sensitive to the sequence~\cite{wartell}.
It is known that this process is not only determined by the fraction of strong 
(GC) or weak (AT) bonds. The sequence specific order is also important. 
Specific sequences can reveal a high opening rate despite a high fraction of 
GC base-pairs~\cite{Dornberger}.  Besides the already mentioned UV absorbance experiments,
many ingenious techniques have been devised to study the denaturation
process and the  statistical and dynamical properties of DNA bubbles in general. 
For instance,
Raman vibrational spectroscopy~\cite{urabe,movileanu}, neutron scattering~\cite{grimm}, 
fluorescent correlated spectroscopy~\cite{altan}, and S1-nuclease 
cleavage~\cite{ChoiNuc2004} have recently
put forward as promising experimental tools to gain insight 
in the complex mechanism of DNA denaturation.

In general, despite this significant progress, the experimental 
techniques reveal only indirect information. Hence, 
complementary computational and theoretical studies
are often a requisite to
complete the interpretation of experimental data.
This is, however, difficult due to the astronomical large 
number of atoms that are needed to describe  solvated DNA. 
Besides the number of atoms of DNA itself, 
a sufficiently large number of water molecules and counter ions should be included.
Any full-atom approach is henceforth limited 
to very short DNA sequences and, for the longest sequences that can be
studied, meaningful bubble statistics cannot be obtained.
This has created need for mesoscopic theoretical
models that allow to study  long DNA sequences of hundreds or even thousands
of base-pairs~\cite{azbel,slucia,poland,kittel,PB,PBD}.
While most of these models try to mimic
the system by an Ising-like model,
the Peyrard-Bishop-Dauxois model~\cite{PB,PBD} (PBD) relies 
on a continuous approach using an effective force-field as function of the
base-pair separation.  
Although more complicated than the Ising type models, 
the PBD model has the advantage that it can describe the DNA sequence 
in a more detailed manner than just a simple array of open and closed states
and it allows to study dynamics 
as well.
An important essence of the PBD 
model is the nonlinear stacking interaction which 
reproduces the experimentally measured sharp phase transition
of long homopolymers~\cite{PBD}. 
Moreover, the model, parameterized for heterogeneous 
DNA chains, has given accurate results for denaturation curves of 
short heterogeneous DNA
sequences~\cite{CAGI}.
Although the PBD model is a very strong simplification of the
actual DNA molecule in solution, 
the qualitative and even quantitative agreement 
with numerous experimental findings have given confidence to this  
model and to its theoretical results for which yet no direct experimental
information is available.   

It were these findings that inspired Choi \emph{et al.}~\cite{ChoiNuc2004,KalosEPL}
to compare the signal of S1 nuclease cleavage experiments
with the formation of bubbles of a certain size obtained
from molecular dynamics (MD) simulations of PBD model. 
The detection of bubbles at a certain size requires
the identification of configurations that contain series of consecutive open 
base-pairs which is very difficult to accomplish experimentally.  
Still, as argued in~\cite{ChoiNuc2004}, the S1 nuclease 
enzymes can selectively cleave the large temporary openings while leaving
the smaller openings intact, hindered by their own physical size.
The amount of cleavages at certain positions in the DNA chain results
in a signal that becomes visible after a certain time of incubation (about
45 min.~\cite{ChoiNuc2004}). The obtained S1 nuclease signal showed a 
remarkable correspondence with the calculated probability profile 
for bubbles containing ten or more base-pairs from the MD simulations of 
the PBD model~\cite{ChoiNuc2004}.   
Moreover,
both experimental and theoretical graphs showed 
clear dominant peaks around the Transcription Start Site (TSS) where
the biological transcription is initiated. 
A similar result had been reported by 
Benham \emph{et al.}~\cite{PNAS1,benham2,benham3,benham4,benham5} who also found a connection between bubble formation and regulatory loci using a theoretical model. 
However, there are two crucial differences between the work of Benham \emph{et al.} and Choi 
\emph{et al.} 
First, the methodology of Benham \emph{et al.}
is specified to detect very large openings upto 100 base-pairs 
in kilobase sequences, while
the work of Choi \emph{et al.} investigates much smaller openings $\sim 10$ in sequences 
of the order  $\sim 100$ base-pairs. The second and most important difference
is that work of Benham studies the bubbles \emph{in vivo} which includes 
torsional effects that are generated by other molecules. The apparent 
evidence of Choi~\cite{ChoiNuc2004} suggested that 
spontaneous bubbles \emph{in vitro} already bear the signature of biological 
activity.  
A remarkable result that was summarized by the 
statement: \emph{DNA directs its own transcription}~\cite{ChoiNuc2004}.

Unfortunately, 
this statement had to be reconsidered  due to  more accurate results  
by us~\cite{vanerpPRL}
using a direct integration method that is orders-of-magnitude
faster than MD. An important difficulty with MD or Monte Carlo
is that large bubbles appear only
seldom so that the
statistical significance can be questioned even for very long 
simulation periods.
Our accurate results did not support the previously found 
results at some crucial points. As in~\cite{ChoiNuc2004}, they indicated that bubbles 
might appear more
easily in the biological active sites due to its higher content of  
AT as compared to a random sequence. However, contrary to~\cite{ChoiNuc2004}, the most dominant
peak did  not appear at the TSS for the sequences under study nor did the promoter sequences
have a much higher opening profile as compared to biologically inactive sequences.
Hence, the statistical information on bubbles obtained by the PBD model was found to be  
insufficient to make very accurate predictions on transcription start sites
or to discriminate between promoter sequences and biologically inactive 
sequences as was suggested before~\cite{ChoiNuc2004,KalosEPL}.
This leaves open the following possibilities:
(i) either the transcription sites cannot be predicted by
the information on thermally induced bubbles  alone but
require more complex interactions including, for instance, 
superhelical stress, or (ii)
the bubble hypothesis of Choi \emph{et al.} still  
holds, as suggested by the S1 nuclease experiments,
but a more accurate theoretical model is needed to support these findings.

The main subject of this article is to give a detailed description of the direct integration method introduced in~\cite{vanerpPRL} and to show some examples of the calculated bubble statistics for some biologically active and inactive sequences. We will also investigate the validity of the PBD model by
applying
this method 
to calculate quantities that allow a more  direct comparison
with experiments. 
This article is organized as follows: we will first give a short introduction to the PBD model in 
Sec.~\ref{secmod},
followed by a theoretical discussion on what we will call the double stranded 
DNA ensemble (dsDNAE) in 
Sec.~\ref{secint}.  
The latter is needed to give meaningful results when applying the PBD model to finite chains.
Then, in Sec.~\ref{secdenbub}, we give some important definitions concerning the bubble statistics
of DNA expressed in microscopic terms such  that it can be calculated by computer experiments.
In Sec.~\ref{secmet} we introduce the direct integration method including all the technicalities involved.
This derivation results in an algorithm that implies a repetitive numerical integration scheme using
a Newton-Cotes rule. The efficiency of several Newton-Cotes schemes, such as 
rectangular, trapezoidal, Simpson's $\frac{1}{3}$-rule, Boole's rule, and 11-point Newton-Cotes rule,
are examined and compared in Appendix~\ref{secnc}. In Sec.~\ref{secbub} 
we show some numerical results of the bubble probability profiles of  a biologically active promoter sequence and 
two artificial Fibonacci sequences. We confirm the previous findings:
there is no enhanced opening at transcription start sites or at promoter sequences in comparison to biologically inactive sites and sequences  that have a similar (local) AT content. Then, in Sec.~\ref{secden} we investigate the 
validity of the PBD model using the direct integration method to calculate $l$-denaturation curves
which can be measured experimentally by the recently introduced quenching 
technique\cite{MontEPL,MontJMB,MontPRL}.
These results clearly indicate that some essential ingredients are
missing in the present PBD model. This implies that the PBD model
should be improved
and that the bubble hypothesis of Choi \emph{et al.} could still hold when an 'ideal theoretical model'
is considered.
In Sec.~\ref{seccon}, we end with a general discussion and make some suggestion
that could lead to an improved theoretical model.

\section{The PBD model}
\label{secmod}
The PBD model reduces the myriad degrees of freedom of DNA
to a one-dimensional chain of effective atom compounds 
describing the relative base-pair 
separations $y_k$ from the ground state positions.
The total potential energy $U$ for an $N$ base-pair DNA chain is then given by 
\begin{eqnarray}
U(y^N) \equiv V_1(y_1)+\sum_{k=2}^N V_k(y_k) +  W(y_k,y_{k-1}).
\end{eqnarray}
Here, $y^N\equiv \{y_k \}$ denotes the set of relative base pair positions and
$V_k$ and $W$ are the two PBD-potential energy functions given by
\begin{eqnarray}
V_k(y_k) &=& D_k \Big( e^{-a_k y_k}-1\Big)^2 \label{PBpot} \\
W(y_k,y_{k-1}) &=& \frac{1}{2} K \Big( 1+\rho e^{-\alpha(y_k+y_{k-1})}\Big)(y_k
- y_{k-1})^2 \nonumber
\end{eqnarray}
The first term $V_k$ is the on site Morse potential describing the 
hydrogen bond interaction between bases on opposite strands. 
$D_k$ and $a_k$ determine the depth and width of the Morse potential  
and are different for the weak AT and strong GC base-pair.
The stacking potential $W$ 
consists of a harmonic and a nonlinear term. 
An important reason for the success of this model lies in the $\rho$-term
which was introduced in~\cite{PBD} as an improvement upon the original 
Peyrard-Bishop (PB) model~\cite{PB}.  This original PB model  
can be retrieved by taking $\rho=0$. 
The precise analytical shape of $W(y_k,y_{k-1})$ in Eq.~(\ref{PBpot}) is 
not crucial. What is important is that
for $\rho>0$, the effective coupling constant of the stacking interaction
drops from $K'=K(1+\rho)$ down to $K'=K$ whenever either $y_k$ or $y_{k-1}$
becomes significant larger than $\alpha^{-1}$.
It is thanks to this additional term that 
the observed
sharp phase transition in denaturation
experiments~\cite{Inman}
can be reproduced.
It is important to note the $+$ sign in Eq.~(\ref{PBpot}). This makes 
the stacking potential $W(y_k,y_{k-1})$ not a simple 
function of the relative
distance $|y_k-y_{k-1}|$. It was found that,
after replacing $e^{-\alpha(y_k+y_{k-1})}$ with 
$e^{-\alpha|y_k-y_{k-1}|}$ in Eq.~(\ref{PBpot}),  the denaturation
transition becomes continuous again as in the 
original PB model~\cite{peydau96}.
However, Eq.~(\ref{PBpot}) is surely not the only possible possible 
potential that can reproduce the sharp transition.
Recently, an alternative potential  $W(y_k,y_{k-1})$ was suggested 
in~\cite{joyeux} which also seems to generate a sharp denaturation 
and only depends $|y_k-y_{k-1}|$. 
This shows that reproducing experimental curves alone is definitely not enough
to uniquely determine the effective potentials. Interpretation of the 
physical mechanism that lead to the sharp denaturation transition 
is a prerequisite
for the justification of the effective models. 
The discussion of this mechanism is definitely not completely settled,
but the argumentation that relies in the PBD model seems very plausible,
as the $\rho$-term mimics the effect
of decreasing overlap between  $\pi$
electrons when one of two neighboring base move out of stack.

After modeling homogeneous DNA, Campa and Giansanti generalized
the PBD model for the heterogeneous case~\cite{CAGI,CAGI2}.
The in total 7 parameters 
$K=0.025$ eV/\AA$^2$, $\rho=2$,
$\alpha=0.35$~\AA$^{-1}$, $D_{w}=0.05$ eV, $D_{s}=0.075$ eV, 
$a_{w}=4.2$~\AA$^{-1}$, $a_{s}=6.9$~\AA$^{-1}$, 
were derived by fitting to experimental  
denaturation curves of short heterogeneous  DNA segments. 
The subscripts $w$ and $s$ refer to the type of base-pair  at site $k$
in Eq.~(\ref{PBpot}). Here, $D_w$ and $a_w$ are used for the weak AT 
base-pairs and  $D_s$ and $a_s$ are used for the strong GC  
base-pairs. The ratio between $D_w$ and $D_s$ reflects the ratio
between the number of hydrogen bonds forming the AT and GC base-pair bonding.
In fact, the reason to  fix this ratio is not really justified as 
the depth of the Morse potential 
does not only reflect the hydrogen bond linking (which is in the order
of 0.2 eV per hydrogen bond), but also the repulsive 
interactions of the phosphate groups and the effect of the solvent.
Still, the absolute and relative magnitude of the 
effective weak and strong interactions
seem to be more or less correct as
this parameterization could reproduce the experimental denaturation curves
of several short DNA sequences as tested in ~\cite{CAGI,CAGI2}.

Despite these accomplishments, 
it is also important to realize the limitations of the model. 
The PBD model treats the A and T bases and the G and C bases as 
identical objects. The stacking interaction $W(y_k,y_{k-1})$ is also 
independent of the nature of the bases at site $k$ and $k-1$.
Experimental measurements~\cite{Breslauer1982,Breslauer1986,SantaLucia1996} and 
theoretical calculations~\cite{Ornstein78,Saenger,pierre,hobza1,hobza2} have shown 
that these are rather crude approximations. 
Future work might aim to improve upon this.

\section{The $\textrm{\bf ds}$DNA ensemble}
\label{secint}
In this section we will assert the need of special ensemble
that we will call the double stranded DNA ensemble (dsDNAE) and
we will give its mathematical definition. 
The reason that we will not use the full NVT or NVE ensemble is because
the results based on the PBD model have not much meaning in these ensembles
whenever finite DNA chains are considered.
The original papers using the PBD model were all performed in the 
thermodynamic limit of an infinite DNA chain where this problem does not appear.
It is in this limit that one can show, 
using the transfer integral technique~\cite{NONLIN},
that 
the uniform PBD-DNA sequence undergoes a very sharp 
phase transition~\cite{theod} upon heating, which is first order
except in a cross over region near the transition temperature that is
so narrow that it is not accessible to experiments.
The difficulty of finite sequences is that PBD model basically represents
a single DNA chain in an infinite solution. 
Hence, whenever the dsDNA completely separates, the two strands are free
to go to very large separations without cost of energy due to the plateau
of  the Morse potential. 
In  experiments,  where the amount of solvated DNA is not 
infinitely diluted,  this effect is counterbalanced by the hybridization 
mechanism where two single stranded chains in solution come together and match 
their complementary bases. This implies that, per definition,
the PBD model cannot reproduce the experimental data, 
which are based on finite concentrations,
using equilibrium statistics
in the full phase space. 
A confinement of the phase space is
\emph{always} necessary.
These can be done hiddenly using a series of reasonable short 
MD~\cite{PBD,dauPRE,joyeux} or Monte Carlo~\cite{ares} 
simulations starting  from a certain distribution of initial configurations.
Here, the finite simulation length prohibit the  
boundless exploration of the completely separated states. 
However, this strategy will naturally generate results that 
depend on the choice of initial conditions and the simulation length
which is not completely under control especially at temperatures
near the melting transition~\cite{PBD,dauPRE}. 
Alternatively, 
one could restrict configuration space 
by adding an infinite wall
such that $y_k$ for all $k$ cannot exceed a certain maximum value~\cite{ZZLC} 
or by adding a small positive slope
to the plateau of  the Morse potential~\cite{theod}. 
These approaches 
still allow for complete 
denaturation and recombination of the two strands, but prevent 
separations 
of very large distances. 
This recombination, however, is quite artificial
as the one-dimensional model does not 
allow for misfolding, the creation of
bulge-loops~\cite{neher} or the recombination with a different 
strand in solution. 
Therefore, we chose to focus to these configurations only that belong to the 
dsDNAE that we will introduce here.
In microscopic terms, a configuration $\{y_k\}$ is called a double stranded DNA (dsDNA) molecule
when $y_k < \xi$ for at least one $k \in [1:N]$ with $\xi$ the opening
threshold definition.
Similarly, a configuration is completely denaturated whenever
$y_k > \xi$ for all $k$. All configurations  assigned as dsDNA together with their 
corresponding  Boltzmann-weight comprise the dsDNAE.

The statistical average of a certain function $A(y^N)$ in the full phase space 
is standardly defined as
\begin{eqnarray}
\lo A \rc \equiv  \frac{ \int \mathrm{d}y^N A(y^N) \varrho(y^N)}{
\int \mathrm{d}y^N  \varrho(y^N)}
\end{eqnarray}
with $\mathrm{d}y^N \equiv \mathrm{d}y_N
\mathrm{d}y_{N-1} \ldots  \mathrm{d}y_1$, $\varrho=e^{-\beta U}$ the
probability distribution density, and $\beta=1/k_B T$ with $T$ the temperature and $k_B$ the Boltzmann constant. In order to define the ensemble average in the
dsDNAE we introduce following characteristic functions that indicate whether a
certain base-pair is open or closed.
\begin{eqnarray}
\theta_k(y_k)\equiv \theta(y_k-\xi), \qquad  \bar{\theta}_k(y_k) \equiv \theta(\xi-y_k)
\label{def_thetas}
\end{eqnarray}
Here $\theta(\cdot)$ equals the Heaviside step function. $\theta_k$ equals 1
if the base-pair is open and is zero otherwise.
$\bar{\theta}_k$ is the reverse.
Now, the ensemble average of $A(y^N)$ in dsDNAE can be expressed
as a weighted average using the weight function $\mu(y^N)$: 
\begin{eqnarray}
\lo  A(y^N) \rc_\mu &\equiv& \frac{\lo A(y^N) \mu \rc }{\lo \mu \rc} \label{def_wa}
\end{eqnarray}
with
\begin{eqnarray}
\mu  &\equiv & 1-\prod_{k=1}^N \theta_k \label{Eqmu}
\end{eqnarray}
 To shorten the notation we have dropped the $y_k$
dependencies.
In Eq.~(\ref{Eqmu}), $\mu=1$ except when all bases are open; then $\mu=0$.
The dsDNAE removes all difficulties
concerning the unnormalizability of the full phase  space equilibrium distribution. 
Besides the opening threshold definition $\xi$, it does not add any 
new (hidden) parameters to the PBD model as in previous examples. 
At temperatures sufficiently below the denaturation transition, the dsDNAE
gives a good representation of the actual experimental situation where
only a fraction of the DNA is in the single stranded state. It is reasonably simple
to use MD in the dsDNAE using a biasing-potential, e.g.~\cite{vanerpPRL}
\begin{eqnarray}
\label{Vbias}
V^{\rm bias}(y_{\rm min})= 
\left\{ \begin{array}{ll}
(y_{\rm min}-\xi)^6 & \textrm{ if } y_{\rm min} > \xi \\0 &
\textrm{ otherwise }
\end{array}\right. \\
\textrm{with } y_{\rm min}=\textrm{MIN}[\{y_k\}] \nonumber
\end{eqnarray}
This bias yields an additional force to the system that is always zero
except when the dsDNA is at the point of complete denaturation. Then 
it gives a strong repulsion to the last closed base to prevent the 
complete opening of the
whole molecule. 
Although, 
MD is certainly much less efficient than the direct integration method expressed in
Sec.~\ref{secmet}, MD using the biasing force~(\ref{Vbias}) can still be useful 
for calculating properties that do not allow the factorization necessary for the 
integration method or dynamical properties.
At higher temperatures, the contributions of single stranded DNA, to e.g. UV absorbance, 
can no longer be neglected. Luckily, recent experimental techniques allow to 
selectively subtract the contributions of the single stranded molecules to the signal~\cite{MontEPL,MontJMB,MontPRL} such that,
effectively, the dsDNA signal can be obtained.  Hence, also at higher temperatures,
the theoretical PBD calculations using dsDNAE can be compared with experimental results.

\begin{figure}[ht!]
\begin{center}
\includegraphics[width=6cm,angle=-90]{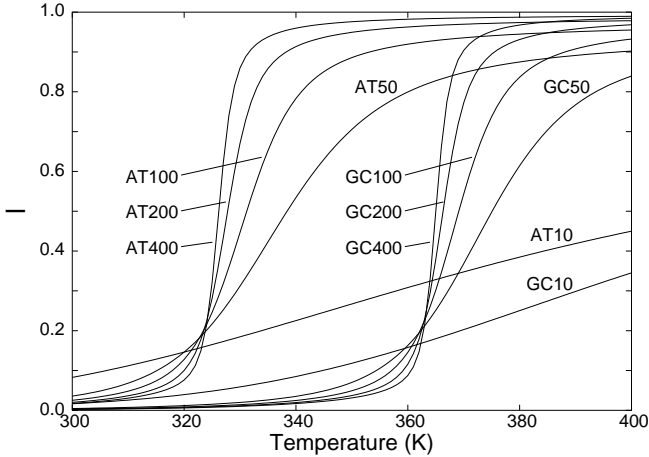}
\end{center}
\caption{$l$ (see Eq.~(\ref{def_l})) as function of  temperature for homogeneous
 AT and GC chains of
different lengths. One can clearly see that, when the length increases,  the cur
ves resemble more
and more a  sharp step function. For all sequences free boundary conditions were
 applied. }
\label{figtrans}
\end{figure}

It is an interesting mathematical problem 
why the complete separation does not
disturb the thermodynamic case. 
In fact, this can be understood invoking one-dimensional random walk theory.
This reveals that, for a fixed configuration of the infinite DNA chain, 
one should always meet a closed 
base-pair when making a walk in one direction along  the chain~\footnote{This
is a result from P\'olya~\cite{Polya} who proved that  one- and two-dimensional 
random walks always return to their origin. In fact, any site will be visited after an
infinite number of steps. The probability to
return after an infinite number of steps to the origin is associated with the 
P\'olya's number.  This number is 1 for one- and two-dimensional systems, but less than 
one for higher dimensions.}. 
Hence, $\mu$  is always 1 for the infinite case and,
thus, the infinite chain remains in the dsDNAE at all times.
It is important to note that, therefore, the constraint to 
keep always one base-pair closed,
does not destroy the phase transition.  On contrary,
the additional constraint allows to study thermodynamic limit 
using finite approximants in a much more controlled way.
Fig.~\ref{figtrans} shows the denaturation curves of finite homopolymers  
of increasing length.
The results was obtained by the direct integration method of Sec.~\ref{secmet},
but could as well been obtained using MD with the bias potential~(\ref{Vbias}).
The denaturation curves of the 400 GC and 400 AT base-pair sequences resemble
already closely  the discontinuous step function, that would result from an infinite chain, and allow
to estimate  the denaturation temperatures quite accurately.  
In contrast,
previous analysis using MD  without any bias had much more difficulty to determine
the denaturation temperature due to huge variations in the melting region despite
the use of very long sequences upto
16384 base-pairs~\cite{dauPRE}.

\section{denaturation curves and bubble probability matrices}
\label{secdenbub}
Using the definitions of~\cite{MontEPL,MontJMB,MontPRL}  
we can call $f$ the fraction of open base-pairs and  $p$ the fraction
of open molecules. With the use of Eqs.~(\ref{def_thetas}) we can give the following mathematical
expressions  
\begin{eqnarray}
f &=& \frac{1}{N} \sum_{k=1}^N \lo \theta_k \rc \nonumber \\
p &=& \lo \prod_{k=1}^N \theta_k  \rc   \label{def_fp}
\end{eqnarray}
Moreover, we introduce
$l$~\cite{MontEPL,MontJMB,MontPRL} as the fraction of open base-pairs provided that the molecules is in the  double stranded state
\begin{eqnarray}
l &=&   \frac{1}{N} \sum_{k=1}^N \lo \theta_k \rc_\mu \label{def_l}.
\end{eqnarray}
Eq.~(\ref{def_fp},\ref{def_l})  obey the relation $l=(f-p)/(1-p)$. 
The quantity $l$ is sometimes called the average fractional bubble length.
However, this is not completely true as more than one bubble
may occur simultaneously in the same sequence.
For the infinite case,
we have $f=l$ and $p=0$ as explained in Sec.~\ref{secint}.
However,
we cannot reproduce the experimental $f(T)$ and $p(T)$ curves for finite chains
as we have, strictly speaking, $f(T)=p(T)=1$ at all temperatures
in the PBD model.
Therefore, we will focus on the behavior of $l(T)$ which can be measured
by the quenching technique of Zocchi and 
co-workers~\cite{MontEPL,MontJMB,MontPRL}. Indirectly,
$f(T)$ could be obtained from $l(T)$ using the phenomenological approach
of Campa and Giansanti~\cite{CAGI}. This approach, however,
requires two additional parameters that have to be fitted to experiments.
Therefore, we believe that the calculation of $l(T)$
gives the most
direct 
comparison with experimental data.

Of course, the experimental UV absorbance signal 
cannot be literally related to the fraction of open base-pairs as it is not
a binary type  measurement that detects whether the base is open or closed. 
Moreover, the theoretical definition of 'open' and 'close' is a bit ambiguous as it depends 
on the choice of opening threshold $\xi$.
Still, it is known that the UV absorbance  changes 
quite abruptly when bases move out of stack, which validates the
$\theta$-like expressions~(\ref{def_fp}) and (\ref{def_l}).
Moreover, it was found that, at least, the qualitative aspects of the 
theoretical denaturation curve are not too sensitive to $\xi$ when chosen
within a reasonable  interval   ($\sim$ [1 \AA: 2 \AA] ) .
Nevertheless, the theoretical definitions (\ref{def_fp}) and (\ref{def_l}) are, 
not the only ones proposed in literature. 
In Refs.~\cite{ares,buyu} another functional form of Eq.~(\ref{def_fp})
was used
\begin{eqnarray}
f' &=& \frac{1}{N} \sum_{k=1}^N \theta\big(\lo y_k \rc-\xi \big)  
\label{fwrong}
\end{eqnarray}
We believe, however, that Eq.~(\ref{fwrong}) should be considered 
as imprecise as
the UV signal is almost a binary indication of the stacking state
of a base pair and, hence, cannot be 
related to the mean position $\lo y_k \rc$ of the bases.
 
Besides denaturation curves, 
the statistical method introduced in Ref.~\cite{vanerpPRL} allows
to study bubbles of a given size. 
The importance to study bubbles of a given size 
was suggested by Choi \emph{et al.}~\cite{ChoiNuc2004,KalosEPL} 
as its signal could be related 
to S1 nuclease cleavage experiments and possibly could tell more about
its biological function than the mean $\lo y_k \rc$ 
or the probability of opening $\lo \theta(y_k-\xi) \rc$. 
Before giving the definition of, what we call, the bubble probability matrix,
we will need to introduce the following auxiliary function:
\begin{eqnarray}
\theta_k^{[m]} &\equiv& \bar{\theta}_{k-\frac{m}{2}}  
\bar{\theta}_{k+\frac{m}{2
}+1} \prod_{k'=k-\frac{m}{2}+1}^{k+\frac{m}{2}} \theta_{k'} \textrm{ for $m$ even} 
\nonumber \\
&\equiv& \bar{\theta}_{k-\frac{m+1}{2}}  
\bar{\theta}_{k+\frac{m+1}{2}} \prod_{k'
=k-\frac{m-1}{2}}^{k+\frac{m-1}{2}} \theta_{k'} \textrm{ for $m$ odd}
\end{eqnarray}
which is 1 (0 otherwise) if and only if $k$ is at the center of a bubble that 
has exactly size $m$. 
For even numbers it is a bit arbitrary where to place the center, 
but we defined it as the base
directly to the left of the midpoint of the bubble. 
The bubble probability matrix $P_{\rm bub}(k,m)$  is, now, 
defined as the probability
to have  
a bubble of size $m$ centered at base-pair $k$ provided that the molecule 
is part of the dsDNAE. Hence,
\begin{eqnarray}
P_{\rm bub}(k,m) \equiv \lo \theta_k^{[m]}\rc_\mu  \label{def_Pbub}
\end{eqnarray}
In principle,
$P_{\rm bub}(k,m)$ contains all the information on the bubble statistics in
a DNA sequence. Still, it is useful to calculate other quantities as well.
From physical and biological perspective, it might be useful to know
the ability to participate in bubbles. Therefore, we introduce
the $P_{\rm part}(k,m)$ probability which is the probability to participate
in a bubble  of at least $m$ sites.
\begin{eqnarray}
P_{\rm part}(k,m) &\equiv& \sum_{m' \geq m}^{\{ m' \textrm{: even}\}} \, \, \,
\sum_{k'=k-m'/2}^{k+m'/2-1} P_{\rm bub}(k',m') \nonumber \\
&+& \sum_{m' \geq m}^{\{ m' \textrm{: odd}\}} \, \, \,
\sum_{k'=k-(m'-1)/2}^{k+(m'-1)/2} P_{\rm bub}(k',m') \label{def_Ppart}
\end{eqnarray}
This quantity is less mathematically stringent as
it is independent of where you assign the position  
of the bubble.  Note that this quantity is still somewhat different from the
projection in Ref.~\cite{vanerpPRL} where each bubble is still associated to
one base-pair position only. In variance with $P_{\rm bub}(k,1)$,
the bubble participation probability  $P_{\rm part}(k,1)$ is directly related
to the simple opening. Hence, $P_{\rm part}(k,1)= \lo \theta_k \rc_\mu \neq  
P_{\rm bub}(k,1)$.

\section{The direct numerical integration method}
\label{secmet}
The two quantities $\lo \theta_k \rc_\mu$ and 
 $\lo \theta_k^{[m]} \rc_\mu$ that appear in Eq.~(\ref{def_l}) and 
(\ref{def_Pbub}) 
can be expressed using partition function integrals:
\begin{eqnarray}
\lo \theta_k \rc_\mu = \frac{Z_{\theta_k}-Z_\Pi}{Z-Z_\Pi} \nonumber \\
\lo \theta_k^{[m]} \rc_\mu = \frac{Z_{\theta_k^{[m]}}}{Z-Z_\Pi},
\label{eqZs}
\end{eqnarray}
which are defined by:
\begin{eqnarray}
Z &=& \int  {\mathrm d} y^N e^{-\beta U(y^N)}  \nonumber \\
Z_{\theta_k} &=& \int {\mathrm d} y^N e^{-\beta U(y^N)}
\theta_k \nonumber \\
Z_{\theta_k^{[m]}} &=& \int {\mathrm d} y^N e^{-\beta U(y^N)}
\theta_k^{[m]} \nonumber \\
Z_\Pi &=&  \int  {\mathrm d} y^N e^{-\beta U(y^N)} \times \prod_j
\theta_j. \label{def_Zs} 
\end{eqnarray}
In Eq.~(\ref{eqZs}),
we used the fact that $(\theta_k)^2=\theta_k$ and $\theta_k \bar{\theta}_k=0$. 
Note that  $Z$, $Z_{\theta_k}$, and $Z_\Pi$ are infinite, but the differences 
$Z-Z_\Pi$ and $Z_{\theta_k}-Z_\Pi$ are finite and well defined.

Now, as all integrals $Z_X$ are of the 
factorizable form
$Z_X=\int \mathrm{d} y^N a_X^{(N)}(y_N,y_{N-1})
\ldots a_X^{(3)}(y_{3},y_{2}) a_X^{(2)}(y_2,y_1)$
we can use following iterative scheme to determine the $Z_X$ integrals:
\begin{eqnarray}
z^{(2)}_X(y_2) &=& \int  {\mathrm d} y_1 \, a_X^{(2)}(y_2,y_{1}) \nonumber \\
z^{(3)}_X(y_3) &=& \int  {\mathrm d} y_2 \, a_X^{(3)}(y_3,y_{2}) z^{(2)}_X(y_2) 
\nonumber \\
\ldots && \nonumber \\
z^{(N)}_X(y_N)&=&  \int  {\mathrm d} y_{N-1} \, a_X^{(N)}(y_N,y_{N-1}) 
z_X^{(N-1)}
(y_{N-1}) \nonumber \\
Z_X&=&  \int  {\mathrm d} y_{N} \, z_X^{(N)}(y_{N}). 
\label{sucint}
\end{eqnarray}

The calculation of $z_X^{(k)}(y_k)$ for a discrete set of $n_{\rm grid}$
values $y_k$ 
requires only $n_{\rm grid}^2$ function 
evaluations whenever $z_X^{(k-1)}$ is known. Hence, a total 
of $N \cdot n_{\rm grid}^2$ function evaluations are required
instead of $n_{\rm grid}^N$ which is a huge improvement.  

An alternative technique was introduced in
Ref.~\cite{ZZLC} where the
$a_X^{(k)}(y_k,y_{k-1})$ kernels are expanded into  a proper basis-sets.  
After this expansion,
the integrals, like in Eqs.~(\ref{def_Zs}), turn into simple matrix 
multiplications
which can be evaluated efficiently.
It was found that performance of such a method depends strongly of the right
choice of basis-set functions. The implementation of this method is, 
therefore, probably 
a bit more
involved than the direct integration scheme of Eq.~(\ref{sucint}). 
Most likely, this method will be more efficient to 
calculate quantities as $\lo y_k \rc$ 
that are written as averages of continuous functions, than, for instance,
$\lo \theta_k \rc$ which involves a discontinuous step-function.
The latter would require a much larger expansion when using continuous
basis-set functions.

The factorization of $Z_X$ into $a_X^{(k)}$ kernels is generally not unique.
Our choice for  $a^{(k)}$ for the partition function $Z$ is the following
\begin{eqnarray}
a^{(k)}(y_k,y_{k-1}) =
\left\{ \begin{array}{ll}
e^{-\beta [W(y_k,y_{k-1})+V_{k-1}(y_{k-1})] }
& \textrm{if }   k \neq N  \\
e^{-\beta [W(y_k,y_{k-1})+V_{k-1}(y_{k-1})
+V_{N}(y_{k}] } &  \textrm{if }   k = N
\end{array} \right. \label{akernel}
\end{eqnarray}
and for $a_{\Pi}^{(k)}$ and $a_{\theta_q}^{(k)}$
\begin{eqnarray}
a_{\Pi}^{(k)}(y_k,y_{k-1}) &=&
a^{(k)}(y_k,y_{k-1}) \theta_k(y_k) \theta_{k-1}(y_{k-1})  \\
a^{(k)}_{\theta_q}
(y_k,y_{k-1})& =&
\left\{ \begin{array}{ll}
a^{(k)}(y_k,y_{k-1}) & \textrm{if } k \neq q,q+1\\
a^{(k)}(y_k,y_{k-1}) \theta_k(y_k) & \textrm{if } k = q\\
a^{(k)}(y_k,y_{k-1}) \theta_{k-1}(y_{k-1}) & \textrm{if } k =  q+1
\end{array} \right. \nonumber
\end{eqnarray}
where we use again that $\theta_{k}^2=\theta_{k}$. Similar expressions 
can be derived for $a^{(k)}_{\theta_q^{[m]}}$. 

In order to perform the numerical calculation,
we need to  define some proper cut-offs where we can stop the integration. 
It is  natural to stop the 
integration whenever the weight of a certain configuration
$\varrho=e^{-\beta U(y^N)}$ drops below a certain threshold value $\epsilon$.
It is clear that the energy diverges and, hence, $\varrho$ vanishes  whenever for a certain $k$
the position $y_k$ takes a very large negative value or when the relative 
distance $|y_k-y_{k-1}|$ becomes very large.
To be in safe limits, we calculate the integration cut-offs for
the pure AT-chain. If we set the integration boundaries 
such that outside this domain we have $\varrho < \epsilon$
for this sequence, it will also hold for the
pure GC or heterogeneous chain.
The lower limit $L$ of $y_k$ results from 
\begin{eqnarray}
e^{-\beta V_w(L)} < \epsilon \Rightarrow
L \lesssim -\frac{1}{a_w}\ln \Big[ \
\sqrt{ \frac{|\ln \epsilon|}{\beta D_{w}} }+1 \Big]
\label{def_L}
\end{eqnarray}
To define the maximal distance $d$ between two neighbors 
we assume 
that  $\rho e^{-\alpha(y_k+y_{k-1})}$ is almost zero. This yields
\begin{eqnarray}
e^{-\beta \frac{1}{2} K d^2 } < \epsilon \Rightarrow d \gtrsim 
\sqrt{\frac{2 |\ln \epsilon|}{\beta K}}.
\label{def_d}
\end{eqnarray}
If $|y_k -y_{k-1}|$ exceeds the value $d$ at any $k$, the probability distribution 
$\varrho(y^N)$ 
must have
decreased below the threshold  $\epsilon$ so that we can stop the integration.
The upper limit 
$R$ is obtained as follows.
Again neglecting the anharmonic $\rho$-term,
the configuration with the lowest stacking energy $\sum W(y_k,y_{k-1})$ and  
with a maximal total stretch $|y_N-y_1|=S$ is obtained whenever 
equidistant positions are taken such that $|y_k-y_{k-1}| = S/(n-1)$. Then, the total stacking energy
equals $(N-1) \frac{1}{2} K (S/(N-1))^2 = \frac{1}{2} K S^2/(N-1) <
 \frac{1}{2} K S^2/N$. Therefore, the maximum displacement of each base, for configurations 
that belong to a double stranded configuration, and with $\varrho(y^N) > \epsilon$, cannot exceed
$R$ given by
\begin{eqnarray}
R &\gtrsim& \xi+S  \textrm{ with $S$ defined by } 
e^{-\beta \frac{1}{2} K S^2/N } = 
\epsilon \nonumber \\
\Rightarrow R &\gtrsim& \xi+\sqrt{N} d.
\label{def_R}
\end{eqnarray}
This completes the set of cut-off values. In principle, the cut-off $d$ is not strictly necessary as
$L$ and $R$ are sufficient to start a numerical approach. However, the cut-off $d$ is useful as it
decreases the computational expense considerably. To summarize,
via Eq.~(\ref{def_L}-\ref{def_R}) we have defined three cut-off values
which restrict the configuration space to $L \leq y_k \leq R$ and $|y_k-y_{k-1}| \leq d$ for all $k$.
Any configuration outside this domain must have a Boltzmann weight $\varrho$ below $\epsilon$ and
can, hence, be neglected for the numerical integration.

The integration boundaries increase only slightly upon decreasing $\epsilon$.
Therefore, we took $\epsilon=10^{-40}$ which is 
much smaller than actually needed for our required accuracy~\cite{vanerpPRL}.
As we take a discrete grid with spacing $\Delta y$, the values  
$d, L$, and $R$ must be adjusted to this grid. That is, we require that
$I_d \equiv d/\Delta y$, $I_L \equiv (\xi-L)/\Delta y$ and $I_R \equiv (R-\xi)/\Delta y$ 
should all be integer values. There is another restriction for  the allowed 
values of $I_R$ which depends on the specific numerical integration method and 
will be discussed in Appendix~\ref{secnc}. 
Coming back to Eqs.~(\ref{def_Zs}), 
we actually no longer intend to
calculate $Z$, $Z_{\theta_k}$, and $Z_\Pi$, which are infinite,
but $Z(R)$, $Z_{\theta_k}(R)$, and $Z_\Pi(R)$ 
which have a linear dependence as  function of $R$.  
However, the differences $Z(R)-Z_\Pi(R)$ 
$Z_{\theta_k}(R)-Z_\Pi(R)$ converge very rapidly to a constant value
for $R \rightarrow \infty$.

As the same function evaluations are repeated over and over again in this 
integration scheme (\ref{sucint}),
it is efficient to store following values at the start of the algorithm
using two matrices $M^{(w)}$ and $M^{(s)}$ defined as:
\begin{eqnarray}
M_{ij}^{(w/s)} & \equiv & 
\exp(-\beta W(L+i\Delta y,L+(i+j)\Delta y)) 
\nonumber \\
&\times&  \exp(-\beta[V_{w/s}(L+(i+j) \Delta y)  ]) \label{def_M}
\end{eqnarray}
which are basically the values of two possible $a^{(k)}(y_k,y_{k-1})$ 
functions~(\ref{akernel}) on the grid.
Then, by  defining the vector 
\begin{eqnarray}
\chi_X^{(k)} (i) \equiv z_X^{(k)}(L+i \Delta y),
\end{eqnarray}
the basic operation in Eq.~(\ref{sucint})
\begin{eqnarray}
z^{(k)}_X(y_k) = 
\int  {\mathrm d} y_{k-1} \, a_X^{(k)}(y_k,y_{k-1}) z^{(k-1)}_X(y_{k-1})
\nonumber
\end{eqnarray}
can be recast in following numerical operation
\begin{eqnarray}
\chi_X^{(k)} (i)=\Delta y \sum_j f_j M_{ij}^{(k-1)} \chi_X^{(k-1)}(i+j)
\label{eqvec}
\end{eqnarray}
where $M_{ij}^{(k-1)}$ is either $M_{ij}^{(w)}$ or $M_{ij}^{(s)}$
of Eq.~(\ref{def_M})
depending on the type of base-pair $k-1$. 
Of course, like the end kernel
$a^{(N)}(y_N,y_{N-1})$ in Eq.~(\ref{akernel}), the last matrix in Eq.~(\ref{eqvec})  
should include the additional factor $\exp(-\beta V_N(y_N))$.
The vector $f_j$ depends on the 
specific Newton-Cotes integration method. 
An analyses of different Newton-Cotes schemes is given in Appendix~\ref{secnc}.

The algorithm starts by taking the first vector $\chi_X^{(1)}(i)=1$ and, then, iteratively apply Eq.~(\ref{eqvec}). 
In order to obtain the full vector $\chi_X^{(k)} (i)$, we need to perform a loop where $i$ runs either from 0 till $I_L$ , from
$I_L$ till $I_R$, or from $0$ till $I_L+I_R$ depending on whether $X$ allows   
$y_k$ in Eq.~(\ref{sucint}) to  take values over the closed, open, or  full domain, respectively.
At each $i$, we perform an inner loop over $j$.
Also $y_{k-1}$ might take values in the closed, open or full domain and its value   
is assigned  by the integer $i+j$. 
Hence, similar to $i$ we can write that $g \leq i+j \leq h$ where
$g$ can be either 0 or $I_L$ and  $h$ is either 
$I_L$ or $I_L+I_R$. As $j \sim |y_k-y_{k-1}|$ is also restricted by $|j| \leq I_d$,  the inner loop over $j$
runs from ${\rm MAX}[-I_d,g-i]$  till ${\rm MIN}[I_d,h-i]$. After the double loop over $i$ and $j$, we increase $k$ by one and repeat the procedure. 
This basically defines the complete numerical algorithm, but  still
leaves open what one should  take for the vector $f_j$. This is
discussed in Appendix~\ref{secnc} in which we consider different
integration schemes.

\section{bubble probability matrices}
\label{secbub}
Now we have introduced the mathematical definitions concerning the
bubble statistics in Sec.~\ref{secdenbub} and derived the numerical method
to calculate these properties in Sec.~\ref{secmet} and Appendix \ref{secnc}, we will 
apply this method to specific sequences. In Ref.~\cite{vanerpPRL},
we calculated the bubble probability matrix~(\ref{def_Pbub}) for de 
adeno-associated viral P5 promoter (AAVP5) whose sequence is shown below
\begin{eqnarray*}
\textrm{AAVP5: 5'-} 
&&\textrm{GTGGCCATTTAGGGTATATATGGCCG}\\
&&\textrm{AGTGAGCGAGCAGGATCTCC{\underline A}TTTTG}\\
&&\textrm{ACCGCGAAATTTGAACG-3'}.
\end{eqnarray*}
The TSS is shown by an underscore.

\begin{figure}[ht!]
\begin{center}
\includegraphics[width=8cm]{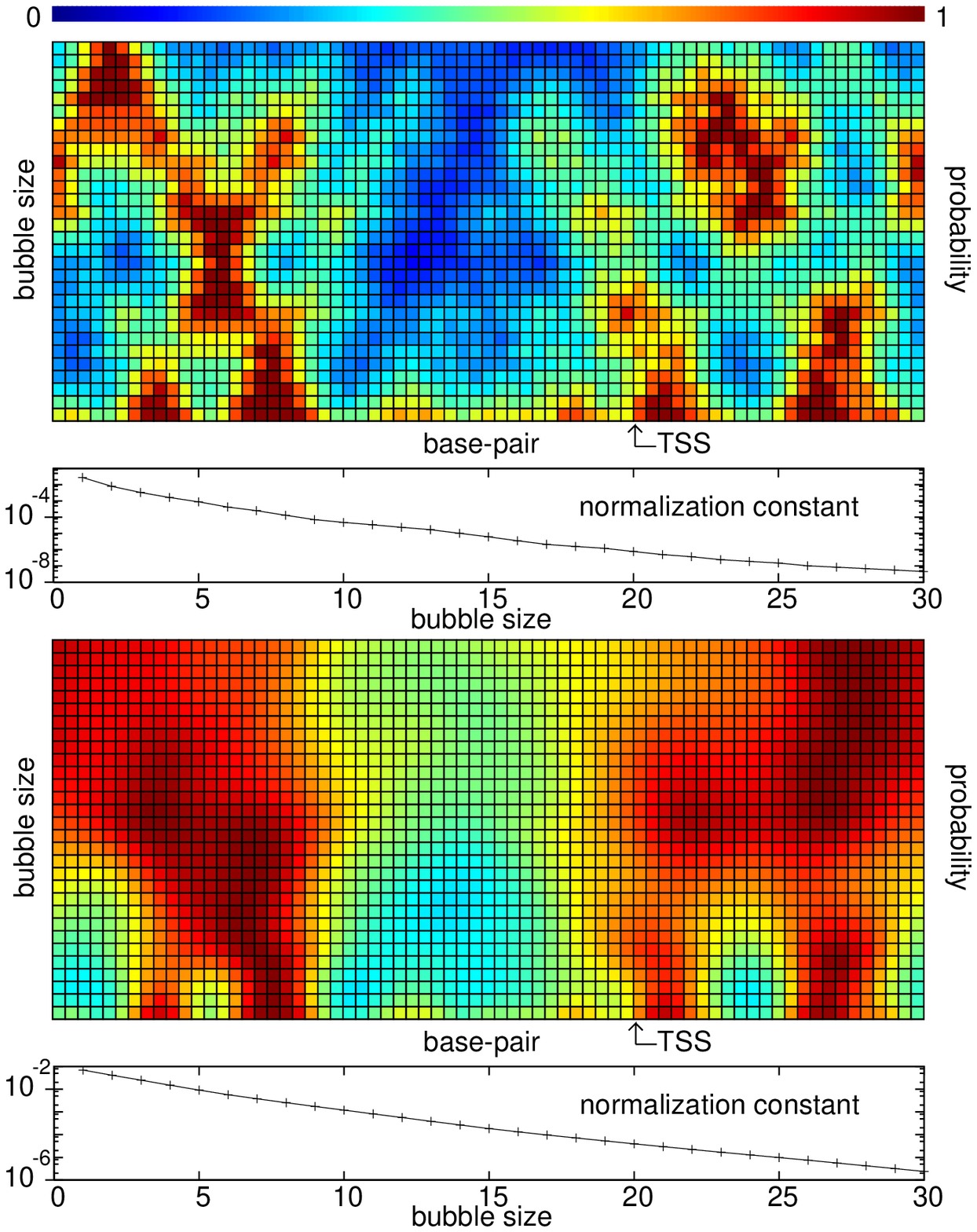}
\end{center}
\caption{(Color online) Bubble statistics matrices for the AAVP5 promoter sequen
ce for $T=300 K$  and openings threshold $\xi=1$ \AA. Top two panels show the bu
bble matrix
$P_{\rm bub}(k,m)$ of Eq.~(\ref{def_Pbub}) and the lower two panels show the bub
ble partition matrices
$P_{\rm part}(k,m)$ of Eq.~(\ref{def_Ppart}). Each row $m$ of the first and thir
d panel is normalized by the maximum value of the matrix at the given bubble siz
e $m$. The normalization constants as function
of $m$ is depicted in the panels below.}
\label{figp5}
\end{figure}

In Fig.~\ref{figp5},
we show the same results as Ref.~\cite{vanerpPRL} 
for a slightly different threshold value ($\xi=1$ 
\AA~ instead of 1.5 \AA ) together with the bubble partition 
matrix~(\ref{def_Ppart}).
As we are not interested in the boundary effects,
we replicated the chain at both ends,
but only computed the statistics for the middle chain.
This eliminates the effects of the free ends, which,
otherwise, would yield very large opening probabilities at boundary sites~\cite{vanerpPRL}.
We calculated the bubble probability matrix~(\ref{def_Pbub}) up 
to bubbles of size $m=50$ (only up to $30$ is shown in Fig.~\ref{figp5}) 
and the bubble partition matrix from 
Eq.~(\ref{def_Ppart}). 
Note that, for reasons of visualization,
we have applied for each row a normalization approach
in these two figures. The normalizing constants, 
which are the maxima in each row, are depicted in the panels below.
Considering these results, one can see that the 
probability for bubbles is approximately exponentially decreasing as function
of the bubble size.
The bubbles of size ten have probabilities of the
order of $\sim 10^{-4}$. This explains the difficulties of previous
MD results~\cite{ChoiNuc2004} as the detection of such a large bubble
is a true rare event on the time-scale accessible by MD.
On the other hand, the numerical integration method allows to obtain 
accurate results for even much larger bubbles. This can be important
for the  study of biological phenomena as, for instance, transcription elongation
involves   
DNA openings that are larger than  ten bases~\cite{Science2}.
The method allows to obtain accuracies of less than one percent error after only a few 
hours of computation which would otherwise take 200 years when using MD~\cite{vanerpPRL}.

Although Fig.~\ref{figp5} shows indeed somewhat enhanced  opening
in  the biologically active regions, it shows that it is certainly not true that
the TSS has a much higher opening probability than the other sites as was found  in the
foregoing less accurate MD results~\cite{ChoiNuc2004}. In fact, the -30
region shows equal probabilities for opening and and even higher probabilities
when bubbles of size $\sim 10$ are considered.
Inspection of the lowest rows in Fig.~\ref{figp5} basically reflects
the AT-rich parts of the sequence. The position
of the preferential opening for the larger bubbles 
can be reasonably understood as a merging effect; 
two small bubbles that are close in distance act as the precursor 
of a larger bubble whose center is in the middle of the two smaller ones.
The $P_{\rm part}$ matrix has considerable less structure, but shows the 
same tendency. 

To investigate whether promoter sites are  special in terms of 
its bubble probability profile, in Refs.~\cite{ChoiNuc2004,vanerpPRL} a
human coding gene, known to be free of any protein interaction 
sites, was examined. The initial results suggested 
that this sequence had much lower probability for bubbles\cite{ChoiNuc2004},
but the direct integration method showed that the ability for bubble formation
was certainly comparable in magnitude to the promoter sequences~\cite{vanerpPRL}.
Here, we study two other artificial non-promoter sequences.  These are
the following two complementary Fibonacci sequences:

\begin{eqnarray*}
\textrm{Fibonacci-1:} &&
\textrm{ACAACACAACAACACAACACAACAA}\\
&&\textrm{CACAACAACACAACACAACAACACA}\\
&&\textrm{ACACAACAACACAACAACACAACAC}\\
&&\textrm{AACAACACAACAAC}
 \end{eqnarray*}
 and
 \begin{eqnarray*}
\textrm{Fibonacci-2:} 
&&\textrm{CACCACACCACCACACCACACCACC}\\
&&\textrm{ACACCACCACACCACACCACCACAC}\\
&&\textrm{CACACCACCACACCACCACACCACA}\\
&&\textrm{CCACCACACCACCA}
\end{eqnarray*}
which have a total length of 89 and a AT content of 
62 \% and 38 \% respectively. 
The choice for Fibonacci has been made to analyze the hypothetical
enhanced opening of biological sequences in comparison with a ``random''
sequence.  However, as a typical random sequence is poorly defined,
one could come up with any sequence and basically ``prove'' what one
wants. Therefore we studied the Fibonacci sequences rather than two
sequences produced by a random number generator. Although those
Fibonacci sequences are far from random, they are sufficiently
disordered and have the advantage that they doe not contain very long
weak or strong regions due consecutive repetitions.  In addition, we
strictly rule out that the possibility we pick by accident a sequence
that is biologically active as well.

\begin{figure}[ht!]
\begin{center}
\includegraphics[width=9cm]{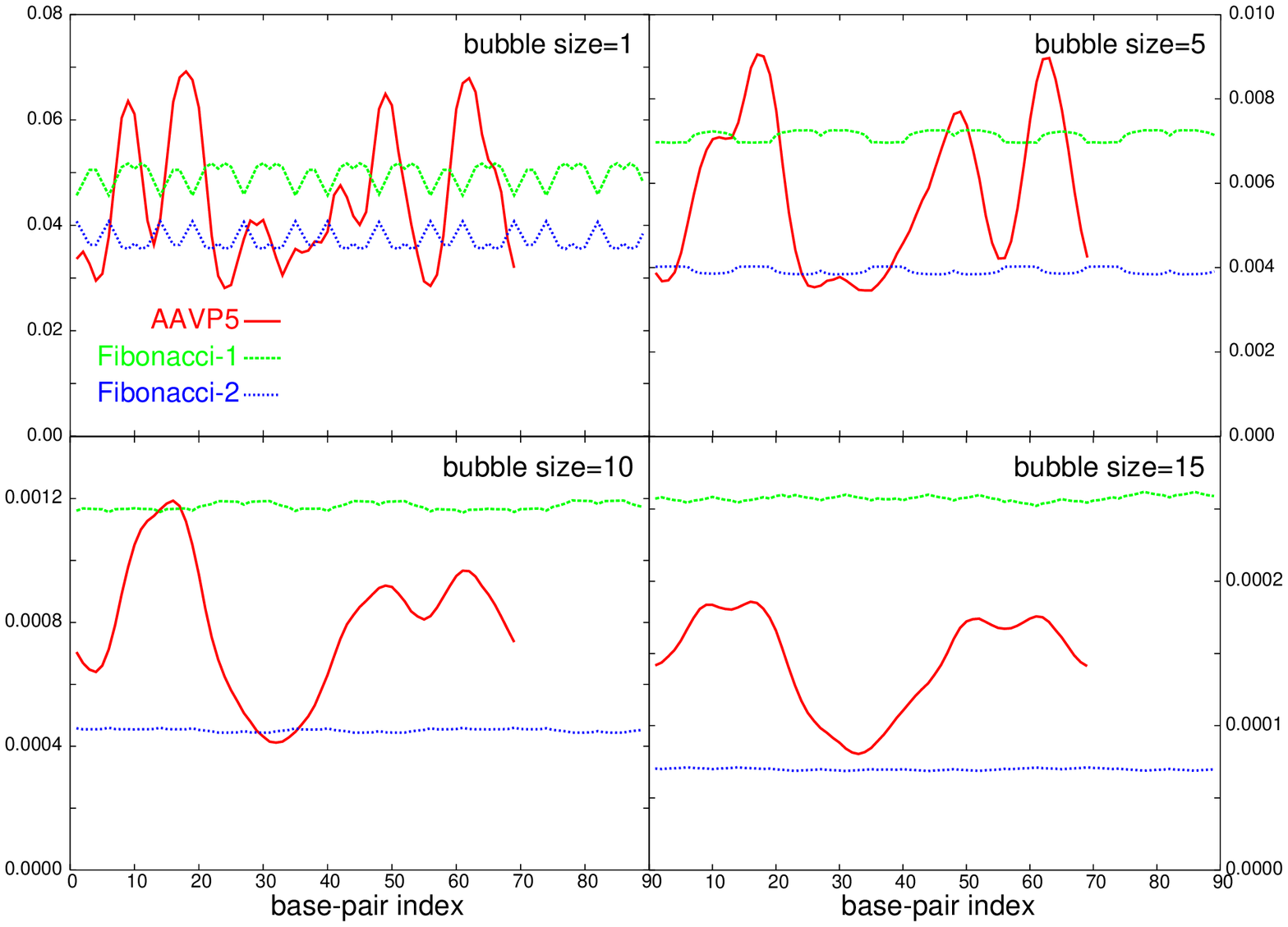}
\end{center}
\caption{(Color online) Bubble statistics, $P_{\rm part}(k,1),  P_{\rm part}(k,5), P_{\rm part}(k,10)$ and $P_{\rm part}(k,15)$, of the AAVP5 promoter and the Fibonacci  sequences.}
\label{figfibos}
\end{figure}

In fig.~\ref{figfibos}, we show the results of $P_{\rm part}(k,1),
P_{\rm part}(k,5), P_{\rm part}(k,10)$ and $P_{\rm part}(k,15)$ for
the Fibonacci sequences together with the results for the AAVP5
promoter. The first panel shows $P_{\rm part}(k,1)$ which equals the
simple opening probability of the individual base in the sequences.
It shows that the promoter sequence has some regions that have a
considerably higher affinity to open up than the Fibonacci sequences.
This is a result of the presence of longer consecutive AT regions in
the AAVP5 promoter. The Fibonacci-1 and Fibonacci-2 sequence have at
most 2 or 1 consecutive weak base-pairs in a row. When we examine
larger bubbles, we see that the base-specific order of the sequences
becomes less important. The extend of the bubble averages out the
effect of the precise order of the weak and strong bases. Hence, the
openings probability profile becomes more and more determined by the
AT content.  This is clearly illustrated by the fact that the promoter
sequence's probability profile for bubbles of size 15 remains strictly
within the two profiles of the Fibonacci sequences at all
sites. Hence, the chance to find a bubble of 15 is at each location
higher in the AT-rich Fibonacci sequence than in the AAVP5 promoter,
despite the absence of long series with consecutive weak AT bases.
This also suggest that the bubble statistics, at least within the PBD
framework, is reasonably predictable by some simple rules based on the
AT content.
Indeed, Rapti \emph{et al.}~\cite{raptiarxiv,raptiarxiv2} suggest that
these PBD bubble profiles could be qualitatively reproduced by
counting the number of AT-bases within a certain window that is a bit
larger than the bubble size considered.
Actual DNA in solution seems to be less predictable on basis of the AT
content alone.  The denaturation steps in long heterogeneous DNA
polymers are very sensitive to the sequence~\cite{wartell} and can
qualitatively change when only one base-pair is changed.
The experimental part of Ref.~\cite{ChoiNuc2004} also suggest that 
actual DNA bubble statistics 
retains strong non-local effects. 
A prerequisite for the understanding of these result would require a
more precise interpretation of the measurements by the S1 nuclease
cleavage technique expressed in microscopic terms. The experimental
signal might well be related to some of the
definitions~(\ref{def_Pbub}) and ~(\ref{def_Ppart}), but probably not
straightforwardly.  Many questions remain such as which range of
bubbles can be detected by S1 nuclease cleavage, where in the bubble
takes the cleavage place, is bubble life-time important, and many
more.  Much more systematic studies are needed.  The results of
Fig.~\ref{figfibos} show that the study of artificial sequences, such
as the Fibonacci sequences, can reveal different structures depending
on the size of bubbles that are detected. Hence, experimental
measurements on artificial periodic and quasiperiodic sequences might
be very useful to give some answers to these intriguing questions.

To summarize this section, our results on the bubble statistics using
the accurate direct integration method do not indicate that
biologically active sites have a stronger thermally induced enhanced
opening than one would expect based on the AT content of the
sequences.  We also examined the effect of higher temperatures upto
350 K and different openings thresholds upto $\xi=2$ \AA. However, the
results remained qualitatively the same. Of course, this does not
necessarily mean that there is not such a relation as this would first
require a validation of the model.  Therefore, in the next section, we
will study the theoretical results of the $l$-denaturation curves that
can give a more direct comparison with experimental data than the
bubble statistics~(\ref{def_Pbub}) and (\ref{def_Ppart}).

\section{Denaturation curves}
\label{secden}
\begin{figure}[ht!]
\begin{center}
\includegraphics[width=8.5cm]{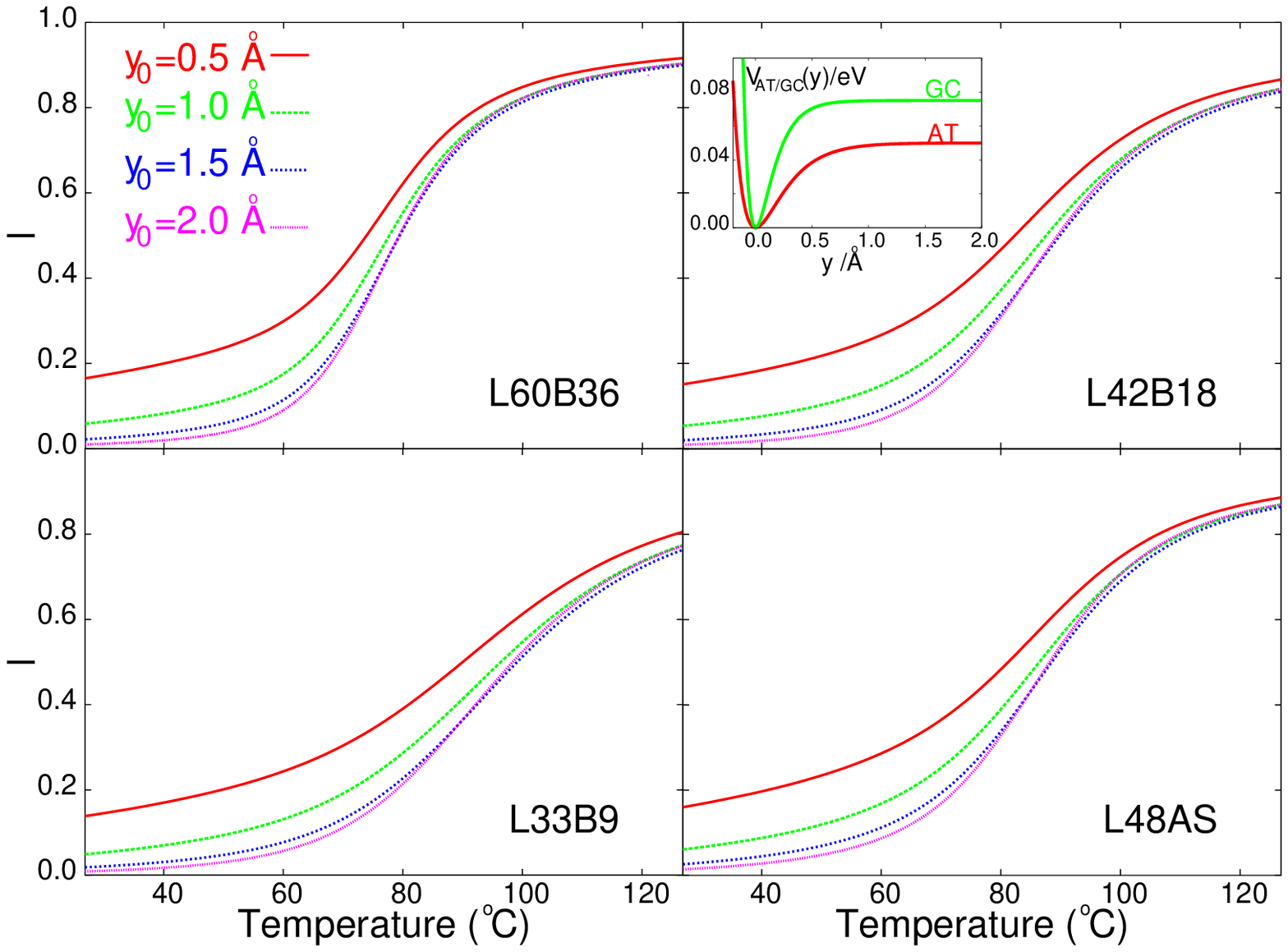}
\end{center}
\caption{$l$ versus temperature for 4 heterogeneous sequences L60B36, L42B18, L33B9, and L48AS
using four different opening threshold definitions $\xi=0.5, 1, 1.5$, and $2$ \AA.
Previous MD results of
Ares \emph{et al.}~\cite{ares} were obtained using $\xi=0.5$ and
the different definition of opening Eq.~(\ref{fwrong}). The inset in the right
upper panel shows the Morse potential $V_k$ of Eq.~(\ref{PBpot}) for
the weak and strong base-pair interaction.
}
\label{figdenat}
\end{figure}

As explained in Sec.~\ref{secint}, the denaturation curves $f(T)$ and
$p(T)$ of Eq.~(\ref{def_fp}) cannot be determined within the PBD
framework. Luckily, the $l(T)$ denaturation curve can be calculated
using the PBD model and can be measured as well using a recently
introduced experimental technique~\cite{MontEPL,MontPRL,MontJMB}. For
several sequences, Montrichok \emph{et
al.}~\cite{MontEPL,MontPRL,MontJMB} reported some anomalous behavior
of $l$ as function of $T$.  These experimental results are, hence, an
excellent benchmark to test the validity of the PBD model.  In this
section, we show the calculated $l(T)$ curves for the L60B36, L42B18,
L33B9, and L48AS given by

\begin{eqnarray*}
\textrm{L60B36:}
&&\textrm{CCGCCAGCGGCGTTATTACATTTAA}\\
&&\textrm{TTCTTAAGTATTATAAGTAATATGGC}\\
&&\textrm{CGCTGCGCC} \\
\textrm{L42B18:}
&&\textrm{CCGCCAGCGGCGTTAATACTTAAGT}\\
&&\textrm{ATTATGGCCGCTGCGCC}\\ 
\textrm{L33B9:}
&&\textrm{CCGCCAGCGGCCTTTACTAAAGGCC}\\
&&\textrm{GCTGCGCC} \\
\textrm{L48AS:}
&&\textrm{CATAATACTTTATATTTAATTGGCG}\\
&&\textrm{GCGCACGGGACCCGTGCGCCGCC}
\end{eqnarray*}
In  Fig.~\ref{figdenat}, we show the calculated
results for these sequence using four values of $\xi$: $0.5, 1.0, 1.5$ and $2.0$ \AA.
It is important to note that the different opening threshold values considered do not change the
qualitative behavior of the  curves. 
The curves with $\xi=2$ \AA~intersect the  lower threshold value curves
$\xi=1$ \AA~ and $\xi=1.5$ \AA. This might seem impossible as each base $k$, that is 
counted as open because $y_k > \xi=2$ \AA, must also be open when a lower opening threshold 
value is considered. However, we should  bear in mind that $\xi$ not only determines
the definition of 'open' and 'closed', but also determines the ensemble via 
Eqs.~(\ref{def_wa},\ref{Eqmu}).  Considering Eq.~(\ref{eqZs}), it is certainly true that 
$Z_{\theta_k}(R)$ is  strictly decreasing as function of $\xi$. However, $Z_\pi(R)$ is 
strictly decreasing as well and, hence, the ratio $[Z_{\theta_k}(R)-Z_{\theta_k}(R)]/[Z(R)- Z_{\theta_k}(R)]$
can actually increase as function of $\xi$.
 
The experimental results for the L60B36 and L42B18 sequences 
contained a remarkable change of slope~\cite{MontEPL,MontPRL,MontJMB}.
This effect could indicate that the melting appears in two steps in which first an AT rich part
of the sequence opens up and is  then followed by a GC rich region in the sequence.  
Our results do not show this signature. This is in contrast with another computational study
by Ares \emph{et al.}~
\cite{ares} which does report some of the experimentally found characteristics. 
However, the  change of slope that they found was  negative 
for the both sequences L60B36 and L42B18, while the
experimental results showed a very sharp positive change of slope in the L42B18 sequence at a temperature of 70 \textcelsius.
Still, the results of Ares \emph{et al.}, for the same sequences we studied, seem to resemble more closely  the experimental results than the ones by us. 
This variance is explained by the following
three reasons: (i) Ares \emph{et al.} used the alternative definition of 'open states' as expressed by 
Eq.~(\ref{fwrong}) instead of Eqs.~[\ref{def_fp},\ref{def_l}].
(ii) They applied a selective use of boundary conditions which were periodic boundary conditions for the sequences
L60B36, L42B18, L33B9 and free boundaries, as in this work,
for the sequence L48AS.
(iii) Ref.~\cite{ares} allowed for complete denaturation as it was based on a series of short MC
simulations without the use of a bias-potential as in Eq.~(\ref{Vbias}).
In fact,
the work \cite{ares} even report on the $f$ and $p$ curves, which
cannot unambiguously be determined as we pointed out in Sec.~\ref{secint}. 
Hence, the deviation from the experimental results  
must imply that the present PBD model is insufficient to reproduce these non-trivial sequence specific order effects.
 
However, the experimental results themselves raise some questions. 
If we were allowed to neglect the
DNA-DNA interaction, the $l(T)$ curve  seems to provide
a signature that is theoretically independent to the concentration of DNA. 
This is exactly why $l(T)$ can be determined 
within the PBD framework.
Still, it would be interesting to verify experimentally whether the $l(T)$ curve is indeed insensitive 
to this concentration. 
Moreover, some of the experimental results are a bit puzzling. 
The experimental $f(T)$ and $p(T)$ denaturation curves  
of the L33B9 sequence, for instance, 
coincide at 75 \textcelsius~ while still $f(75 \textrm{ \textcelsius})=p(75 \textrm{ \textcelsius}) < 1$~\cite{MontJMB}.
As $l=(f-p)/(1-p)$, this
would imply that $l(T)=0$ for $T>75$ \textcelsius.  This finding seems to be unphysical and this is 
probably also the reason that Montrichok \emph{et al.} have depicted the $l(T)$ curve until $T=75$ \textcelsius~ in Ref.~\cite{MontJMB}.  
This indicates that one has to be careful
when translating the UV absorbance experiments in microscopic terms
using Eqs.~(\ref{def_fp},\ref{def_l}). The theoretical development in
this  field would benefit significantly if more 
experimental data based on the quenching technique were available.

\section{Conclusions}
\label{seccon}
The statistics of thermally induced DNA bubbles has become an important subject of theoretical and experimental studies. Besides the fact that it is a interesting subject from a 
purely statistical physics point of view, the relation between thermally induced bubbles and biologically active sites has been subject of recent debate. Mesoscopic models, like the PBD model, are a prerequisite in these studies as the experimental data can usually only give indirect information.
However, even if a good theoretical model is developed, it is not easy to obtain  accurate results
as large bubbles occur only seldom in a microscopic system. 
In previous publications, the inaccuracy inherent to MD have lead to premature conclusions
such that the TSS has a much stronger affinity to form bubbles than any other arbitrary site~\cite{ChoiNuc2004,KalosEPL}.  In a recent publication by us~\cite{vanerpPRL}, we showed,
using a new statistical method that is orders of magnitude faster
than MD,
that this statement had to be reconsidered. 
Although the biologically active sites have some enhanced opening due to their relative 
high content of weak AT base-pairs, the bubble probability profile given by the PBD model
was certainly not sufficient to make accurate predictions on transcription sites or to discriminate between
biologically active and inactive sequences. Hence, this implies that either the biologically active sites
cannot be assigned by the information of thermally induced bubbles alone or the actual PBD model
is insufficient to describe all the sequence specific effects correctly. The S1 nuclease experiments
seem to suggest a correlation between bubbles \emph{in vitro} and transcription sites. It is, however,
not exactly clear how the S1 nuclease measurements  should be translated in microscopic terms that
can be calculated by computer experiments. 

In this article, we have revisited the direct numerical integration technique that was introduced 
in ~\cite{vanerpPRL}. 
We have given a detailed explanation of the algorithm and investigated the performance of different 
integration schemes. Although the higher order Newton-Cotes schemes are better for very 
high precision results with many digits,
the simple Simpson $\frac{1}{3}$-rule or Boole's rule are more efficient if only an accuracy of a few
percent is required. The optimal result is obtained when the Simpson's or Boole's rule
is combined with the simple rectangular rule.  The latter is used when the function vanishes at the 
two integration boundaries. Moreover, we have given a thorough discussion on how 
to treat finite chains using the PBD model by introducing the double stranded DNA ensemble.
This eliminates all the problems due to the unnormalizability of equilibrium distribution in the full space
and gives results that can be compared by experiments performed below the melting temperature.

Within this ensemble, we have defined two types of bubble probabilities. 
$P_{\rm bub}(k,m)$ is the probability that a bubble of exactly size $m$ is centered at base-pair $k$. $P_{\rm part}(k,m)$ is the participation probability that site $k$ is inside a bubble of at least $m$ bases long.  
Our analyses on the AAVP5 promoter sequence and two artificial Fibonacci sequences confirm 
what we found before~\cite{vanerpPRL}. No theoretical evidence was found that 
bubbles appear more frequently 
at transcription sites than at other sites that have a similar AT content.
When larger bubbles are considered, the effect of sequence specific order becomes 
even less important. A recent theoretical study of Rapti \emph{et al.}~\cite{raptiarxiv} confirms this and reveals
that the PBD bubble statistics profile can be qualitatively reproduced by counting the number of AT
within a certain window that is larger than the bubble size. The questions remains whether this is also
true for actual DNA. The S1 nuclease experiments suggest that the behavior of real DNA is more complicated
than that. 

To study the validity of the PBD model, we applied our method to calculate the so-called 
$l$-denaturation curves that 
allow to make a more direct comparison to experimental results. As argued, the standard 
$f$-denaturation curve cannot be obtained without
additional parameters due the problem of normalizability for finite DNA chains.   
Luckily, the $l$-denaturation curves can be measured as well via a recently introduced quenching 
technique~\cite{MontEPL,MontJMB,MontPRL}.
Our theoretical calculations did not reproduce the experimentally found
anomalies of the $l(T)$ denaturation curve.  
This points out a significant weakness of the present PBD model. 
This also implies that the bubble hypothesis postulated by Choi \emph{et al.}~\cite{ChoiNuc2004}
could still be supported by theoretical evidence whenever an 'ideal' DNA model is considered.
The indirect evidence of the S1 nuclease experiments is yet insufficient to make this statement
absolute
as its meaning in terms of microscopic terms is not yet completely understood.
It is also difficult to believe that the statement holds for all TSS as some transcription sites are known 
that consists of at least three consecutive strong base in a row~\cite{galas}. 
More systematic experimental and theoretical studies are required. 

Theoretical improvement
can probably be achieved when a more complicated stacking interaction
is taken into account.
We found that some of the anomalies found by Montrichok \emph{et al.}~\cite{MontEPL,MontJMB,MontPRL} 
could be reproduced  using a different base-pair specific stacking potential 
$W(y_k,y_{-1})$ (\ref{PBpot})~\cite{SantiagoTBP}. However, more complicated potentials might be
needed. It is important to note that the direct integration method is not restricted to the PBD model only.
It can be used whenever the proper factorization (\ref{sucint}) can be applied.  
Our preliminary results indicate that the PBD model could be improved considerably while  still maintaining the one-dimensional character of the model.
This implies that the direct integration method could still be applied for this new class of 
models and will, hence, probably remain an important method for the future theoretical developments in this field.

\begin{acknowledgement}
We would like to thank Johannes-Geert Hagmann for useful discussions and critically reading this manuscript.
TSvE has been supported 
by a Marie Curie Intra-European Fellowships
(MEIF-T-2003-501976) within
the 6th European Community Framework Programme and by the Belgium 
IAP-network. 
SCL has been supported by the Spanish Ministry of Science and
Education (FPU-AP2002-3492), project BFM 2002-00113 DGES and DGA (Spain) and by
the CAI-Europa XXI program. 
\end{acknowledgement}

\appendix
\section{Newton-Cotes integration schemes}
\label{secnc}
In Sec.\ref{secmet}, we
have given the derivation of the direct integration method upto
the numerical implementation which basically comprises an iterative
operation of Eq.~(\ref{eqvec}).
The vector $f_j$  
depends on choice of Newton-Cotes integration scheme. In general, the Newton-Cotes numerical 
integration approximates 
any integral over an finite range 
$\int_a^b g(x) \, {\mathrm d} x$  by
$\Delta y \sum_{i=0}^n f_i \, g(a+i \Delta y)$ with $n=(b-a)/\Delta y$.
From the various Newton-Cotes schemes, we  will discuss the 
simple rectangular rule, Simpson $\frac{1}{3}$-rule,
Boole's rule, and the 11-point Newton-Cotes formula.
The corresponding $f_i$ vectors are listed below.\\
\\
\noindent
Rectangular rule:  
\begin{eqnarray} \label{REC}
f_i = 1 \textrm{ for all } i,
\end{eqnarray}
Trapezoidal rule:
\begin{eqnarray} \label{2NC}
f_i= \left\{ \begin{array}{ll}
\frac{1}{2} & \textrm{ for } i=0,n \\
1 & \textrm{ for } i =1,2,3,\ldots, n-1
\end{array} \right. ,
\end{eqnarray}
Simpson's $\frac{1}{3}$ rule:
\begin{eqnarray} \label{3NC}
f_i= \frac{1}{3} \times \left \{ \begin{array}{ll}
1 &\textrm{ for } i=0,n \\
4 &\textrm{ for } i=1,3,5,\ldots, n-1 \\
2 &\textrm{ for } i=2,4,6,\ldots ,n-2
\end{array} \right. ,
\end{eqnarray}
Boole's rule~\cite{Boole}: 
\begin{eqnarray} \label{5NC}
f_i=
 \frac{2}{45} \times \left\{ \begin{array}{ll}
7 &\textrm{ for } i=0,n \\
32 &\textrm{ for } i=1,5,9,\ldots,n-1 \\
12 &\textrm{ for } i=2,6,10,\ldots, n-2\\
32 &\textrm{ for } i=3,7,11,\ldots,n-3 \\
14 & \textrm{ for } i=4,8,12,\ldots,n-4
\end{array} \right. ,
\end{eqnarray}
and the 11-point Newton-Cotes rule~\cite{AbrSte1972}:
\begin{eqnarray} \label{11NC}
f_i= \frac{5}{299376} \times \left \{ \begin{array}{ll}
16067 &\textrm{ for } i=0,n \\
106300 &\textrm{ for } i=1,11,21,\ldots,n-1 \\
-48525 &\textrm{ for } i=2,12,22,\ldots, n-2\\
272400 &\textrm{ for } i=3,13,23,\ldots,n-3 \\
-260550 & \textrm{ for } i=4,14,24,\ldots,n-4 \\
427368 & \textrm{ for } i=5,15,25,\ldots,n-5 \\
-260550 & \textrm{ for } i=6,16,26,\ldots,n-6 \\
272400 & \textrm{ for } i=7,17,27,\ldots,n-7 \\
-48525 & \textrm{ for } i=8,18,28,\ldots,n-8 \\
106300 & \textrm{ for } i=9,19,29,\ldots,n-9 \\
32134 & \textrm{ for } i=10,20,30,\ldots,n-10 \\
\end{array} \right. .
\end{eqnarray}

The right choice of integration scheme can significantly improve the 
precision of the method. One cannot say in advance that  the highest
order scheme is always preferable. This can depend on the shape of the 
function $g$, the applied integration boundaries, and the required precision.
In order to study the accuracy of the integration methods,
we applied the different schemes (\ref{REC}-\ref{11NC}) on 
the standard integral $\int_{a}^{\infty} e^{-x^2} \, {\mathrm d} x$ 
where we take
$a=-\infty, 0$ and 1.
We take a numerical cut-off such that  $|x| \leq 10$ on the integration domain. 
In general,  the higher order Newton-Cotes numerical integration schemes
require that the total number of integration intervals $n$
must    be multiples of a certain value. These are 2, 4, and 10 for, 
respectively, Simpson's rule, Boole's rule 
and 11-point Newton-Cotes.  
However,  as the function vanishes at 
the right boundary ($x=10$ in our numerical approach) 
we can take the semi-infinite analogue where we start with
$f_0$ at the  point $x=a$ (or $x=-10$ if $a=-\infty$) and then simply 
continue with $f_1, f_2, \ldots $
until the point $x=10$ without requiring the correct ending $f_n=f_0$.

\begin{figure}[ht!]
\begin{center}
\includegraphics[width=8cm]{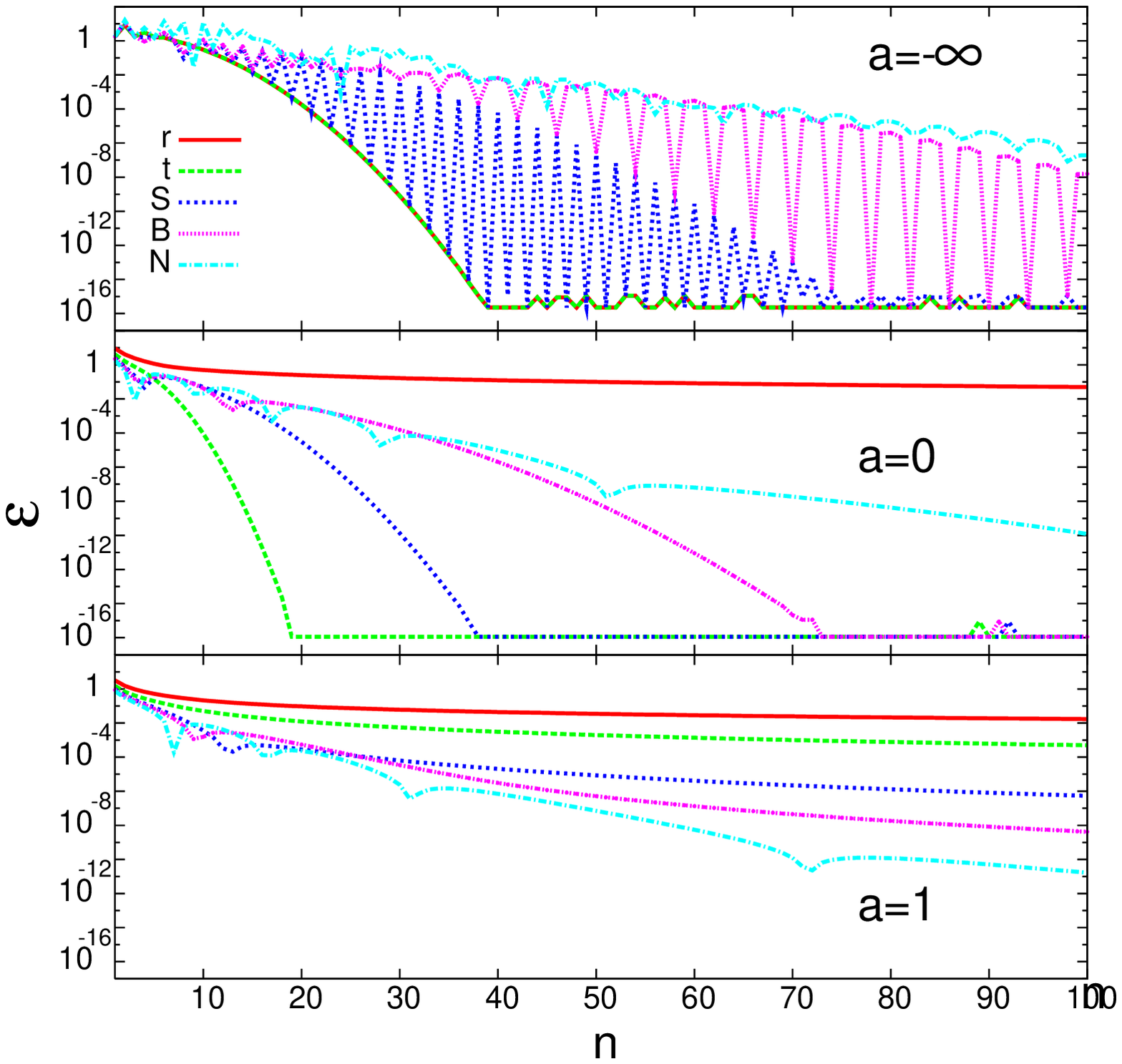}
\end{center}
\caption{(color online) The absolute error for the integration
of $\int_a^\infty  \exp(-x^2) \, \mathrm{d} x $ as function of the number of intervals $n$ in the numerical approach. The rectangular rule (r), trapezoidal rule (t), Simpson's $\frac{1}{3}$-rule (S), Boole's rule (B),
and the 11 point Newton-Cotes scheme (N) are compared. Three cases are considered: $a=-\infty$ (top), $a=0$ (middle), and $a=1$ (bottom). The horizontal plateau in the first two panels is a result
from the cut-off at $\pm 10$.}
\label{figApp}
\end{figure}
In fig.~\ref{figApp}, we have plotted the integration errors as function of $n$
obtained by the five Newton-Cotes methods and the three values of $a$. 
We see that the highest order scheme is not always the best choice. 
In fact, for the integration over the full range ($a=-\infty$), 
the simple rectangular rule is identical to the trapezoidal rule, but
far superior to the other methods~(\ref{3NC}-\ref{11NC}). 
Naturally, as the function vanishes at both ends, 
the result  would not  change much upon shifting the 
initial point to $-10+ \Delta x$. 
Averaging over several shifts 
using Eqs.~(\ref{3NC}-\ref{11NC}) results 
in a weighted summation that approaches the simple rectangular 
rule~(\ref{REC}). The optimum performance of the rectangular rule 
on the infinite domain
is, hence, not surprising. 

The trapezoidal rule gives the optimal result for the case $a=0$. Also this is not too surprising as
the function is symmetric and the trapezoidal rule is exactly half the result of the rectangular rule over the
full domain.
For $a=1$ we find, as expected, that the 
11-point Newton-Cotes method gives the best result. 
However,  
only at large $n$ the difference becomes apparent.

We also analyzed the performance of the different Newton-Cotes  schemes for 
the bubble statistics in the PBD model. 
As a benchmark,
we compared the calculated values of $l$~(\ref{def_l}) at temperature $T=300$ K and threshold 
opening $\xi=1$ \AA~ for
a 10 base-pair long homogeneous AT chain with free boundaries.
Considering previous results, 
we  always applied the rectangular rule for the integrals in (\ref{sucint}) when the integrated
function vanishes at both integration or cut-off boundaries in $y_{k-1}$.  These
are either at $y_{k-1}=L$ or at $y_{k-1}=y_k \pm d$. When the integrated function  only
vanishes  at one end, we applied  the semi-infinite variation of one of the Newton-Cotes  formulas 
[\ref{REC}-\ref{11NC}].
It is important to notice that, as we mentioned before,  the distribution function is not vanishing at $y_{k-1}=R$. Hence, the Newton-Cotes  rule needs to be  applied at this boundary for an 
optimal accuracy; i.~e. we start with $f_0$ at this boundary and continue in the negative direction
$R-\Delta y,  R-2 \Delta y, \ldots$ for the numerical integration. 

The integrals with two non-vanishing boundaries 
appear only for the last integrations $Z_X = \int \mathrm{d} y_N Z_X^{(N)}$ in Eq.~(\ref{sucint}) and  when $y_N$ must be  integrated over the open domain only.
Then,  both at the left boundary $y_N=\xi$ as at the right boundary  $y_N=R$, the function
is not necessarily decayed below $\epsilon$.  This also implies that $R-\xi$ is the only interval
that must be a special multiple of $\Delta y$. This must be an multiple of 2 for Simpson's rule,
4 for Boole's rule and 10 for 11-point Newton-Cotes rule and this gives the restriction to the possible 
integer values that $I_R$ can take. 

After these technical details are taken into account, 
the Newton-Cotes formulas  [\ref{REC}-\ref{11NC}] can be applied to the benchmark system 
and allow to compare the different integration methods.
The results are depicted in table~\ref{tab1}.
\begin{table*}[hb!]
\begin{tabular}{|l|l|l|l|l|l|}
\hline
$\Delta y$ & 0.2 & 0.1 & 0.05 & 0.025 & 0.0125 \\ \hline
r&   1.7059& 1.6112& 1.57187104& 1.5520500628& 1.5422425176 \\
t&   1.5506& 1.5339& 1.53313317& 1.5326609537& 1.5325428897 \\
S&   1.5015& 1.5307& 1.53253332& 1.5325035590& 1.5325035346 \\
B&   1.4887& 1.5333& 1.53258630& 1.5325015790& 1.5325035330 \\
N&   1.1940& 1.5156& 1.53183258& 1.5325009318& 1.5325035341 \\ \hline
\end{tabular}
\caption{Analysis of the accuracy of the Newton Cotes integration scheme. $l (10^{-1})$ for a 10 base-pair homogeneous AT chain is shown for different values of  $\Delta y$. 5 integration schemes are compared: rectangular rule (r), trapezoidal rule (t), Simpson's $\frac{1}{3}$-rule (S), Boole's rule (B), and
11-point Newton-Cotes formula (N).
}
\label{tab1}
\end{table*}
These show that it is certainly beneficial to go beyond the simple rectangular or trapezoidal rule. 
Although, higher order schemes like the 11-point Newton-Cotes are presumably better
at very small values of $\Delta y$ and very high precision, at larger values of  $\Delta y$ the Simpson's and Boole's method give better results. 
The highest precision results with $\Delta y=0.0125$ \AA~ 
show an accuracy of 8 digests for both Simpson, Boole and 11-point 
Newton-Cotes, while the computational expense is less than a minute.
Such a performance is far  beyond any MD or MC method even if 
enhanced sampling is applied~\cite{TV74}.

For our purposes, an accuracy a few percent is enough. 
Therefore, considering the results of Fig.~\ref{figApp} and Table~\ref{tab1},  we have chosen to use
Simpson's rule with
a grid spacing of $\Delta y=0.1$. In the results of Sec.~\ref{secbub} and~\ref{secden},
we have always used these parameters.

\bibliographystyle{prsty}

\end{document}